\documentclass[a4paper,fleqn,usenatbib]{mnras}

\usepackage{mathptmx}

\usepackage[T1]{fontenc}
\usepackage{ae,aecompl}

\usepackage{graphicx}	
\usepackage{amsmath}	
\usepackage{amssymb}	

\usepackage{soul}

\title[Dynamical Lifetimes of Asteroids in Retrograde Orbits]{Dynamical Lifetimes of Asteroids in Retrograde Orbits}

\author[P. Kankiewicz, I. W\l odarczyk]{
Pawe{\l} Kankiewicz,$^{1}$\thanks{E-mail: pawel.kankiewicz@ujk.edu.pl}
Ireneusz W\l odarczyk$^{2}$
\\
$^{1}$Institute of Physics, Astrophysics Division, Jan Kochanowski University, Swietokrzyska 15, 25-406 Kielce, Poland\\
$^{2}$Polish Astronomical Amateur Society, Powstancow Wlkp. 34, 63-708, Rozdra{\.z}ew, Poland\\
}

\date{Accepted XXX. Received YYY; in original form ZZZ}

\pubyear{2016}

\begin{document}
\label{firstpage}
\pagerange{\pageref{firstpage}--\pageref{lastpage}}
\maketitle

\begin{abstract}
The population of known minor bodies in retrograde orbits ($i > 90 \degr$) that are classified as asteroids is still growing. The aim of our study was to estimate the dynamical lifetimes of  these bodies by use of the latest observational data, including astrometry and physical properties. We selected 25 asteroids with the best determined orbital elements. We studied their dynamical evolution in the past and future for $\pm$ 100 My ($\pm$ 1 Gy for three particular cases). 

We first used orbit determination and cloning to produce swarms of test particles. These swarms were then input into long-term numerical integrations and orbital elements were averaged. Next, we collected the available thermal properties of our objects and used them in an enhanced dynamical model with Yarkovsky forces. We also used a gravitational model for comparison. Finally, we estimated the median lifetimes of 25 asteroids.

We found three objects  whose retrograde orbits were stable with a dynamical lifetime $\tau \sim 10 \div 100$ My. A large portion of the objects studied displayed  smaller values of $\tau$ ($\tau \sim 1$ My). In addition, we  studied the possible influence of the Yarkovsky effect on our results. 

We found that the Yarkovsky effect can have a significant influence on the lifetimes of asteroids in retrograde orbits.
Due to the presence of this effect, it is possible that the median lifetimes of these objects  are extended. Additionally, the  changes in orbital elements, caused by Yarkovsky forces, appear to depend on the integration direction.
To explain this more precisely, the same model based on new physical parameters, determined from future observations, will be required.
\end{abstract}



\section{Introduction}

Since the discovery of asteroid (20461) Dioretsa in 1999 (MPC: 48396), the results of systematic observations and analysis show that the list of asteroids in retrograde orbits ($i > 90 \degr$) is steadily increasing. According to the MPC classification, most  of these retrograde asteroids belong to either the Centaur, or TNO population. Under certain circumstances, they may also be classified as either Main Belt Asteroids, or inner planet crossers, e.g. NEAs: 343158 (known as 2009 HC82; an Apollo-class asteroid). Retrograde asteroids do not belong to a specific dynamical group nor family. 

During our previous studies concerning retrograde orbits \citep{KankiewiczWlodarczyk2010,KankiewiczWlodarczyk2014a}, a small fraction of retrograde asteroids have been classified as comets after cometary activity was observed, e.g. NEO 2007 VA85, now classified as comet 333P/LINEAR. The  availability of new observational data has led to many such objects being reclassified as comets. Therefore, the results concerning the dynamical evolution of these reclassified objects require subsequent revisions and corrections.

\subsection{Discovery statistics and opportunities}

Before the discovery of the first retrograde asteroid (Dioretsa), small bodies with large inclinations  ($i > 90 \degr$) were typically classified as comets due to an observed display of cometary activity. The known population of retrograde asteroids is constantly growing (even with the account of numerous cases where these bodies were reclassified as comets). In the past two years, the number of retrograde asteroids listed in the MPC database has risen from 56 to 95 as of June 2016. This population may even be greater in size, however, due to observational limits  inherent in detecting distant objects, an assessment of its actual size could prove to be an even greater challenge.

According to \cite{Brasser2012}, there exists a sub-population of Centaurs that are referred to  as HIHQs (high-inclination, high perihelion).   These Centaurs have large orbital inclinations ($ i > 60 \degr$) and their dynamical evolution  is not dominated by giant planet perturbations. This small fraction of Centaurs, studied in this paper, probably originate from the Oort Cloud. Within this sub-population, the probability of finding retrograde asteroids that exhibit similar HIHQ characteristics is very low,  however, when taking the latest discoveries into consideration, their total number remains an open question.

Due to the existence of numerous meteoroids in retrograde, heliocentric orbits \citep{Greenstreet2012, Borovicka2005}, it is expected that some of these bodies can have their source among the asteroids.  New discoveries and dynamical studies of retrograde NEAs can potentially explain the origin of high-velocity meteoroids in retrograde orbits.

\subsection{Classification and origin}

As mentioned by \cite{Brasser2012}, the MPC and JPL definitions of Centaurs do not directly depend on their dynamical properties. Classification criteria of MPC and JPL do not assign retrograde Centaurs to any particular group.

JPL's Small Body Database (JPL SBDB) classifies the small sub-population of retrograde asteroids as TNOs. In most cases, this refers to unnumbered objects, some of which are listed in Table~\ref{object_list}. In our sample of 25 multi-opposition retrograde asteroids, nine objects are classified in SBDB as TNOs.

According to \cite{Chen2016}, there are two  TNOs currently known that have retrograde orbits: 2008 KV42 \citep{Gladman2009}, which was the first discovered retrograde TNO, and 'Niku', which was recently discovered and found to be linked with 2011 KT19.

Retrograde asteroids compose only a small fraction of all distant objects in our Solar System. This small size of this population leads to the  speculation that dynamical lifetimes of these objects are probably short. This   speculation is based generally on scenarios where the orbital resonances or close approaches are a factor that greatly influences the minor body's orbital evolution.

In the case of Near Earth Asteroids, \cite{Greenstreet2012} presented possible scenarios   for the production of retrograde variants. In many cases, a mean-motion resonance (MMR) between an asteroid and a planet (e.g. the Jupiter 3:1 MMR)  induces highly inclined, retrograde orbits for the object. The presence of such a resonance can directly influence the dynamical lifetimes of objects.
If an object were to be ejected from this resonance, then its median lifetime extends from $10^3$ y to  $10^7-10^8$ y.

In the case of Centaurs and Damocloids, \cite{Morais2013} identified the first objects in retrograde resonance with Jupiter and Saturn. They also described in detail the retrograde resonance capture scenarios for each object respectively. After an update of the observational data on retrograde objects, \cite{Marcos2014}  proceeded to study the dynamical evolution of three large Centaurs: (342842, 2011 MM4 and 2013 LU28). In many cases, it was found that high-order resonances with the Jovian planets had induced chaotic behaviour in these Centaurs' orbital evolution.

Overall, these studies all lead to  suggest that these objects had originated in either the Oort Cloud or an alternative, hypothetical reservoir placed at a distance of hundreds of AU, composed of objects that have highly inclined and high-eccentricity orbits.

\section{Methods}

In our simulations, we considered  only those known objects which were in retrograde orbits, classified as asteroids, and possessing sufficiently long observational arcs. In our first  run, we used the 56 retrograde asteroids available in 2014. After taking into account the latest observations and discoveries, we updated our list to  95 asteroids known as of June 2016. During our calculations, some objects were found to exhibit cometary activity and were thus reclassified as comets (e.g. 2007 VA85). In other cases, asteroids were  reclassified. For example, 2014 JJ57 was first classified as a Main Belt asteroid; after new observations,  it was reclassified as a Centaur. 

Additionally, large observational errors for single-opposition (usually short-arc) objects cause a relatively high dispersion of their clones (Section \ref{data-models}). Due to these limitations, we finally selected  25 numbered and multi-opposition asteroids with sufficiently well-determined orbits. They are listed in Table~\ref{object_list} and were used in our long-term numerical integrations. 

\subsection{Initial data and models}\label{data-models}
The starting orbital elements were computed using data from the MPC.  First, we used the orbit determination software Orbfit \citep{Orbfit, Milani2005} and generated 100 clones along the line of variation (LOV) for each nominal orbit. Next (in the main numerical integration), the clones were  propagated backwards and forwards for $10^8$ yr using two dynamical models: a simplified, gravitational model, ('Grav. model') and a model with thermal forces, ('Yark. model'). For the numerical integration, we used the swift\_rmvsy package \citep{Broz2011}. The model of Yarkovsky forces  is described in more detail by \cite{Vokrouhlicky1998} and \cite{Vokrouhlicky1999}. In the Grav. model, we used the starting orbital elements of 8 perturbing planets from the JPL Ephemerides DE405. 

\subsection{Physical parameters}
For the Yark. model,  a set of physical data for each asteroid  was required in order to calculate the Yarkovsky forces acting on each object. 
These data sets consisted of the following physical parameters: radius ($r$), bulk density ($\rho$), surface density ($\rho_{surface}$), thermal conductivity ($\kappa$), thermal capacity ($C$), albedo ($p_V$), infrared emissivity ($\epsilon$), period ($P$), and spin orientation parameters $\lambda$, $\beta$ (or any equivalent information about pole coordinates: e.g. components of the unit vector of spin). 
In the JPL SBDB, we found only one asteroid having an available value of $P = 200 $ h (65407). For the majority of asteroids, the values of the albedo were still unknown. In most cases, thermal parameters were either not determined or unavailable,  thus approximate values  were used instead (Table~\ref{thermal_table}).
We determined the approximate diameters of the asteroids using the expression ~\citep{Fowler1992}:
\begin{equation}
\label{diameter_formula}
D[km]=\frac{1329}{\sqrt{(p_v)}}10^{-\frac{H}{5}},
\end{equation}
where $H$ denotes the absolute magnitude.

To prepare a simplified model of the Yarkovsky accelerations,  common values among all sets of asteroid parameters were  adopted to describe this model.   
The typical values of $p_V$, $\rho_{bulk/surface}$, $\kappa$, $C$, and $\epsilon$ were adopted from the literature:  the albedo of Centaurs was adopted from \cite{Bauer2013},  NEA parameters from \cite{Nugent2012},  densities from \cite{Carruba2014}, surface properties from \citep{Mueller2007},  thermal properties of Centaurs from \cite{Guilbert2011}, commonly used values of $\kappa$ and $\epsilon$ from \cite{Broz2006}, and the rotations and densities of TNO/Centaurs from \cite{Sheppard2008}. The rotation  periods of NEAs, if not determined,  were marked in Table~\ref{thermal_table} as $f(R)$ and calculated  assuming that a typical NEA  1 km in size has a period of rotation \mbox{$P = 5$ h} and  $P$ values are proportional to diameters ($D$).
 Since pole coordinates were unavailable in all cases, we used a random distribution of  spin axis directions for each set of clones. This allowed for estimating mean values of possible Yarkovsky drift. 

All default values for the thermal parameters are listed in Table~\ref{thermal_table} and were used if they had not been observationally determined for that particular object.  In the case of Main Belt Asteroids (MBA), the set of typical parameters listed in Table~\ref{thermal_table} should be read for reference purposes only, because two retrograde objects classified as MBAs were reclassified during our calculations and assigned to other groups.

\subsection{Orbital integration}
We integrated the swarm of clones (100 clones for each object) using the swift\_rmvsy package backward and forward in time, typically $\pm 10^8$ y. For some asteroids with long lifetimes, we decided to extend the integration time to $\pm 10^9$
y. The average elements of each object were computed at each step of the integration. Additionally, the results were weighted assuming a Gaussian probability density function  and assigning a higher statistical importance to data located closer to the nominal orbit.

\begin{table}
\caption{Default values for thermal parameters used in  the Yark. model  for cases where observational results were lacking. $f(H,p_v)$ is calculated by Eq.~\ref{diameter_formula}. In all cases, random spin axis directions  were  assigned.}
\label{thermal_table}
\footnotesize\addtolength{\tabcolsep}{-3pt}
\begin{tabular}{|ccccccccc|}
\hline
Class. & $D$ &	$\rho$  & $\rho_{surface}$  & $\kappa$ & $C$  & $p_V$ & $\epsilon$ & $P$ \\
\hline
Centaurs & $f(H,p_v)$ & 1125  & 1125  &  0.006  &  760  & 0.08 & 0.9 & 8.4 \\
 TNO    &   &  &   &  &   &  &   &     \\
\hline
MBA & $f(H,p_v)$  & 3000 & 1500  & 0.001  & 680 &  0.10 & 0.95 & 6.0 \\
\hline
NEA & $f(H,p_v)$  & 1200  & 1200  &  0.08  &  500 &   0.14 & 0.9 & $f(R)$ \\
\hline
\multicolumn{9}{c}{Units: $D$ [$m$], $\rho$ [${kg}\;{m^{-3}}$], $\kappa$ [$W\;K^{-1}\;m^{-1}$] , $C$ [$J\;kg^{-1}\;K^{-1}$] , $P$ [$h$] }\\
\end{tabular}
\end{table}

\begin{table}
\caption{The list of  numbered and multi-opposition retrograde asteroids selected for our numerical integrations.}
\label{object_list}
\small\addtolength{\tabcolsep}{-4pt}
\begin{tabular}{|c|c|c|c|c|c|}
\hline
Name             &  $i$[$\degr$] & $e$ &  $a$[AU] & $H$      & Class.\\  
\hline
  (20461) Dioretsa   & 	160.4 &	0.90 &	23.83  & 13.8        & Centaur \\
  (65407) 	     & 	119.0 &	0.95 &	54.67  & 12.3        & TNO \\ 
  (330759) 	     & 	170.3 &	0.57 &	8.12   & 12.8        & Centaur \\
 (336756) 	     & 	140.8 &	0.97 &	319.71 &	10.6 & TNO \\
 (342842) 	     & 	105.0 &	0.44 &	11.63  &	9.3  & Centaur\\
 (343158)(2009 HC82) & 	154.4 &	0.81 &	2.53   &	16.2 & Apollo(NEA)\\
  (434620)(2005 VD)  & 	172.9 &	0.25 &	6.66   &	14.3 & Centaur\\
 (459870) 	     & 	165.6 &	0.41 &	10.95  &	11.9 & Centaur\\
    1999 LE31 	     & 	151.8 &	0.47 &	8.14   &	12.4 & Centaur\\
    2000 HE46 	     & 	158.5 &	0.90 &	23.56  &	14.8 & Centaur\\
    2004 NN8 	     & 	165.5 &	0.98 &	97.08  &	15.3 & TNO\\
    2006 BZ8 	     & 	165.3 &	0.80 &	9.61   &	14.2 & Centaur\\
   2008 KV42 	     & 	103.4 &	0.49 &	41.36  &	8.8  & TNO\\
   2009 QY6 	     & 	137.6 &	0.84 &	12.50  &	14.8 & Centaur\\
   2009 YS6 	     & 	147.8 &	0.92 &	20.26  &	14.2 & Centaur\\
   2010 CG55 	     & 	146.3 &	0.91 &	31.96  &	14.2 & TNO\\
   2010 GW64 	     & 	105.3 &	0.94 &	63.27  &	14.9 & TNO\\
   2010 OR1 	     & 	143.9 &	0.92 &	27.13  &	16.2 & Centaur\\
   2011 MM4 	     & 	100.5 &	0.47 &	21.12  &	9.3  & Centaur\\
   2012 HD2 	     & 	146.9 &	0.96 &	61.53  &	15.3 & TNO\\
   2013 BL76 	     & 	98.6  &	0.99 &	1171.9 &	10.8 & TNO \\
   2013 LD16         & 	154.8 &	0.97 &	78.64  &	16.1 & TNO\\
   2013 LU28 	     & 	125.4 &	0.95 &	171.32 &      8.0    & Centaur\\ 
   2013 NS11 	     & 	130.3 &	0.79 &	12.67  &	13.6 & Centaur\\
   2014 JJ57 	     & 	95.9  &	0.29 &	6.99   &	12.7 & Centaur\\
\hline
\end{tabular}\end{table}

\section{Results}

\subsection{Retrograde orbits: Interesting cases}

The only example of a retrograde NEO asteroid is (343158), formerly known as 2009 HC82. Previously, we studied its dynamics and  Earth impact  probability \citep{KankiewiczWlodarczyk2014a}.  In that work, two retrograde NEAs were mentioned, however, the second  object, 2007 VA85, has  since been reclassified as a comet and thus is excluded from our list.
During our previous integrations  of 900 clones of 2009 HC82 \citep{KankiewiczWlodarczyk2014a}, no remarkable planet  collisions were recorded; however,  one Mars co-orbital solution with low probability ($P \simeq 1.469 \cdot 10^{-4}$) was identified  30 My in the future.

\begin{figure}
	\includegraphics[width=\columnwidth]{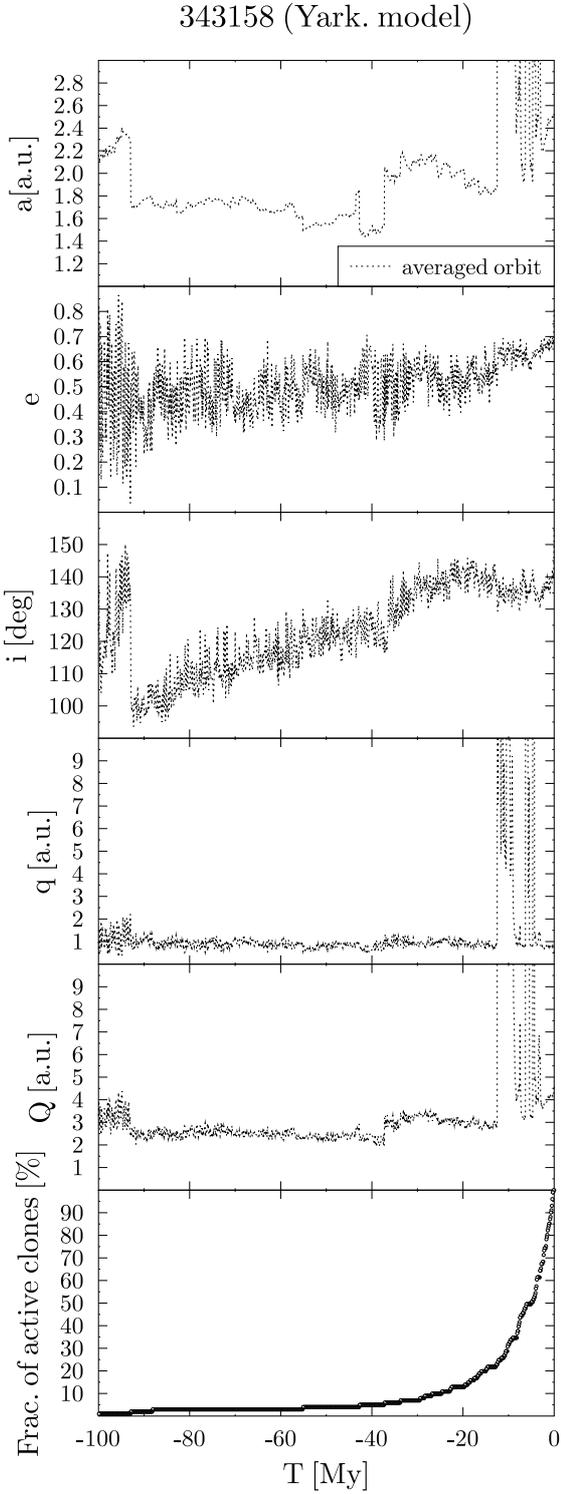}
    \caption{The orbital evolution of retrograde NEO 343158 (2009 HC82), 100 My backward.}
    \label{evo_figure1a}
\end{figure}

\begin{figure}
	\includegraphics[width=\columnwidth]{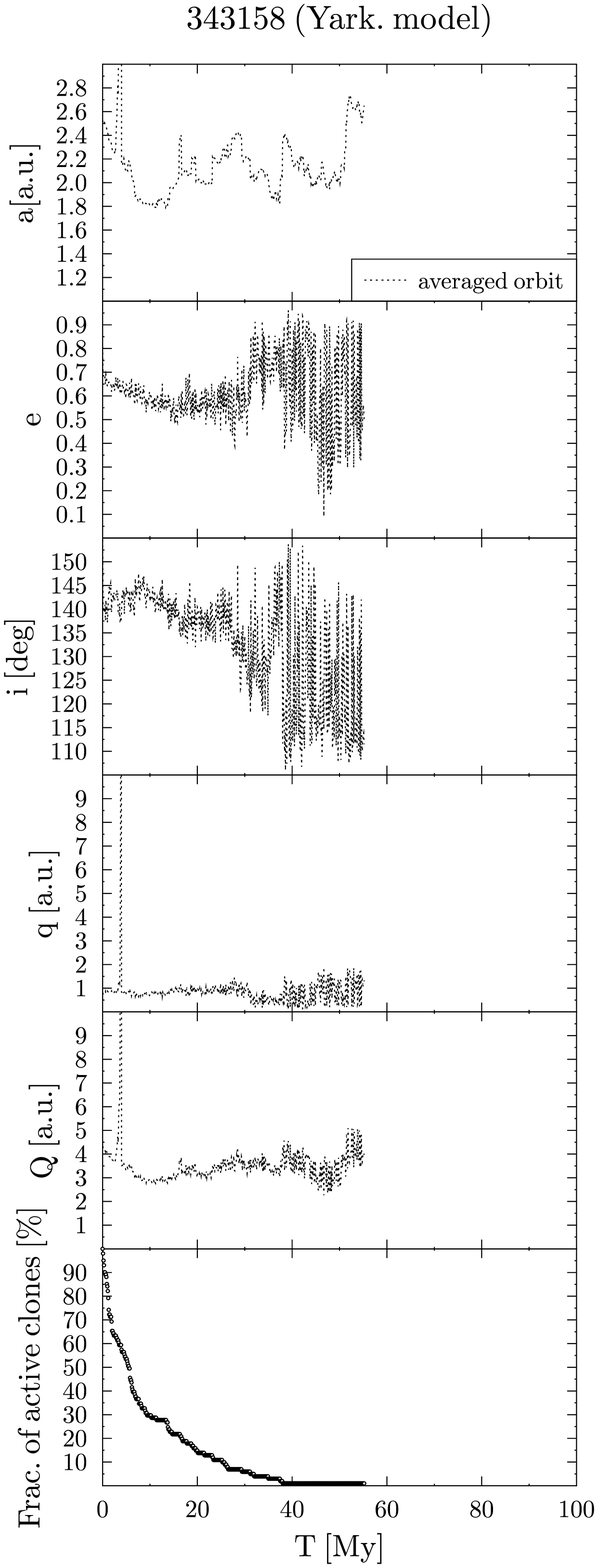}
    \caption{The orbital evolution of retrograde NEO 343158 (2009 HC82), 100 My forward.}
    \label{evo_figure1b}
\end{figure}

During the current numerical integrations,  numerous clones of 343158 were ejected:
 99 \% of them were ejected  during the 100 My backward integration (Fig. \ref{evo_figure1a}). The forward integration yielded an ejection of all test particles onto hyperbolic orbits and none survived longer than 60 My (Fig. \ref{evo_figure1b}).

\begin{figure}
	\includegraphics[width=\columnwidth]{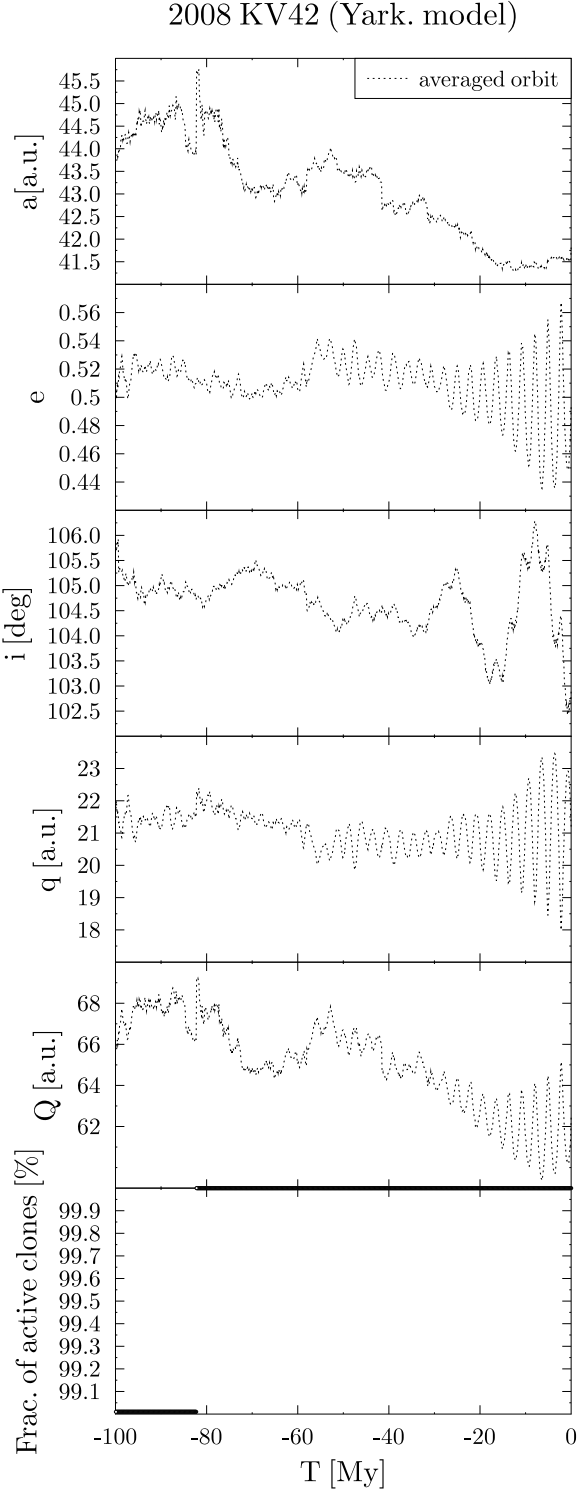}
    \caption{The orbital evolution of 2008 KV42, limited to 100 My backward for comparison with another object (343158, Fig. \ref{evo_figure1a}, \ref{evo_figure1b}). Note that the decay of the clones of this asteroid is relatively small.}
    \label{evo_figure2a}
\end{figure}

\begin{figure}
	\includegraphics[width=\columnwidth]{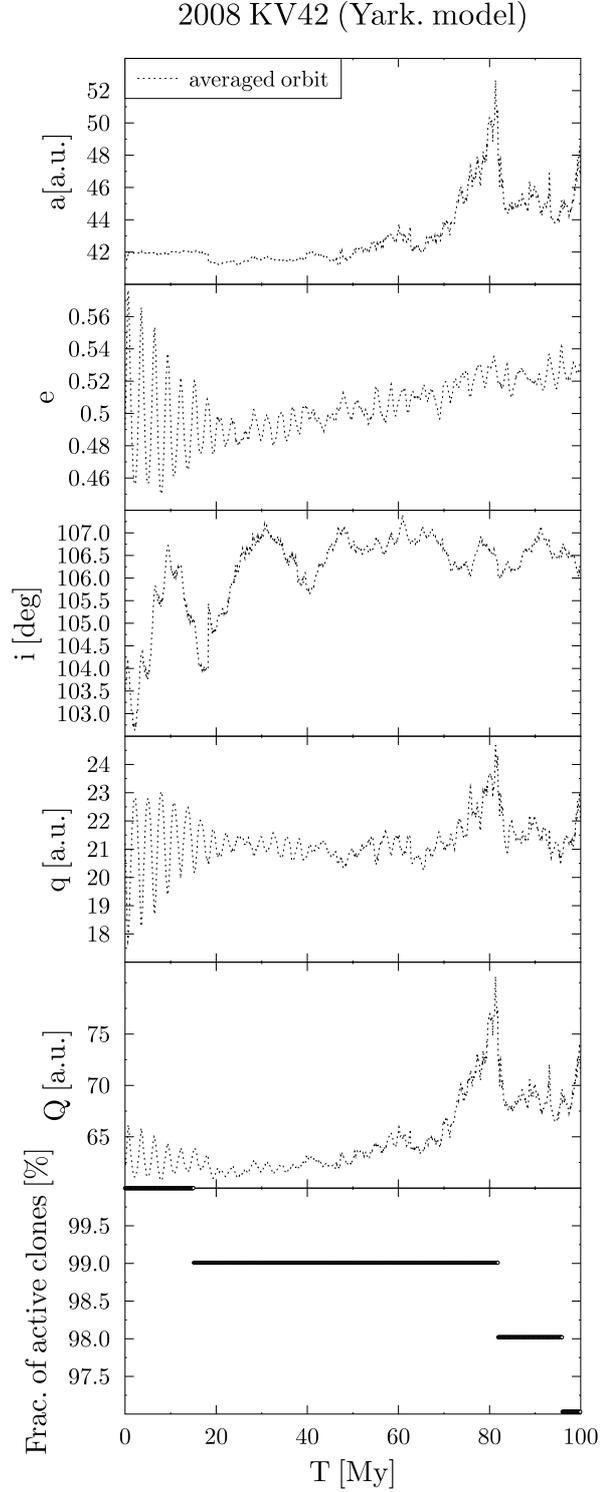}
    \caption{The orbital evolution of 2008 KV42, limited 100 My forward. Most of these clones survived the 100 My integration, but after the extended integration, we obtained 50 \% decay after $\tau$ = 640 My.}
    \label{evo_figure2b}
\end{figure}

The asteroid 2008 KV42 is an example of  an object with a long dynamical lifetime.  For this object, the majority of clones survived the $\pm 100$ My integrations (Figs \ref{evo_figure2a} and \ref{evo_figure2b}). In the case of 2008 KV42, we extended the time of the integration to $\pm 1$ Gy to obtain more information about  its long-time evolution and to  estimate its dynamical lifetime.   An extension of the integration time was needed for (336756) and 2011 MM4, but their lifetimes are shorter, especially in the future (Fig. \ref{lifetime_long2}).

\subsection{Influence of the Yarkovsky effect}

We compared the results obtained by the two dynamical models. This  comparison allowed us to analyze the effect of Yarkovsky forces on long-term evolution. 
In Tables~\ref{yarko_influence_backward} and \ref{yarko_influence_forward}, we show the mean differences in the semimajor axis, eccentricity, and inclination, after 1 My of numerical integration for the two models. The differences in $a$ (where $\Delta a$=$a_{Yark}-a_{grav}$) grow quickly during the integration for several reasons, such as high eccentricity and  the chaotic properties of  trajectories. We suspected that most of our objects are in unstable orbits, especially the retrograde TNOs and Centaurs \citep{Pinilla-Alonso2013}. Due to the fast divergence  of nearby trajectories it is difficult to separate the effect of the Yarkovsky drift from  chaotic behaviour. However, the complexity of the dynamical  model and the application of thermal (i.e. Yarkovsky) forces can be essential in the long-term integration of retrograde orbits. The average change of inclination (${\Delta} i = i_{Yark}-i_{grav}$) tends to be greater in the past (Table~\ref{yarko_influence_backward}) than in the future (Table~\ref{yarko_influence_forward}). The dispersion of ${\Delta} i$ is also greater during backward integration and a similar behaviour is observed for ${\Delta} e$. 

In general, the Yarkovsky force induced a positive drift in  the semimajor axis of prograde rotators and a negative drift in the case of retrograde rotators. It is possible that changes in other orbital elements  such as $e$ and $i$ depend indirectly on this effect as well.  In many cases  in our study, the sign of the eccentricity change depended on the integration direction (Tab. \ref{yarko_influence_backward}, \ref{yarko_influence_forward}). Additionally, the  standard deviation of eccentricity changes was much larger in the backward integration than in the forward integration: $\sigma(\Delta e)_{backward}=0.303$ vs $\sigma(\Delta e)_{forward}=0.0311$.   


There are several probable explanations for the dispersion of elements in the past due to the Yarkovsky effect.
The Yarkovsky drift can potentially cause   migration of  a minor body into  a more chaotic zone. For example, mean-motion resonances (related to  specific values of $a$) or frequent close approaches to planets may cause chaotic behaviour. It is possible that asteroids perturbed by the Yarkovsky force may be more likely to pass through chaotic zones in the past than in the future. This hypothetical mechanism, however, requires stronger evidence supported by a much more accurate dynamical model. In particular, we need more accurate parameters of rotation (like period, direction of rotation, obliquity) to determine the strength and direction of Yarkovsky drift for individual objects.

\begin{table}
\caption{Differences in the mean values of $a$, $e$, $i$  between the Grav. model and Yark. model after -1 My.}
\label{yarko_influence_backward}
\small\addtolength{\tabcolsep}{-2pt}
\begin{center}
\begin{tabular}{|c|r|r|r|}
\hline
Name             &  ${\Delta} a$[AU] & \multicolumn{1}{c}{${\Delta} e$} & ${\Delta} i$ [$\degr$] \\  
\hline
(20461)Dioretsa  & +4.32e+02 &  +0.556 &  -22.861   \\
(65407)          & +1.28e+02 &  +0.426 &  +73.482   \\
(330759)         & +2.69e+02 &  +0.061 &  +13.250   \\
(336756)         & -3.40e+01 &  -0.006 &  +2.956    \\
(342842)         & +8.21e+03 &  +0.224 &  +18.532   \\
(343158)(2009 HC82) & -1.21e-01 &  +0.099 &  -15.102   \\
(459870)(2014 AT28) & +3.17e+03 &  +0.642 &  +23.912   \\
1999 LE31         & +1.81e+02 &  +0.054 &  -0.293    \\
2000 HE46         & -3.71e+02 &  -0.349 &  -20.882   \\
2004 NN8          & -1.87e+02 &  -0.066 &  +26.403   \\
(434620)(2005 VD)   & -1.58e+02 &  -0.417 &  +22.572   \\
2006 BZ8          & +2.06e+01 &  -0.018 &  -25.683   \\
2008 KV42         & -4.32e-01 &  -0.016 &  +1.219    \\
2009 QY6          & -4.56e+02 &  -0.513 &  +0.296    \\
2009 YS6          & -7.45e+01 &  -0.024 &  +52.092   \\
2010 CG55         & -3.68e+01 &  -0.037 &  -16.871   \\
2010 GW64         & -2.02e+02 &  -0.055 &  +11.145   \\
2010 OR1          & -1.03e+01 &  -0.112 &  +6.552    \\
2011 MM4          & -3.75e+02 &  -0.406 &  -11.054   \\
2012 HD2          & -2.60e+02 &  -0.078 &  +6.680    \\
2013 BL76         & -6.07e+04 &  -0.004 &  +19.356   \\
2013 LD16         & +9.44e+01 &  +0.443 &  +24.288   \\
2013 LU28         & -5.26e+02 &  -0.218 &  -23.382   \\
2013 NS11         & +5.80e+03 &  +0.091 &  -5.597    \\
2014 JJ57         & -8.53e+03 &  -0.481 &  -14.656   \\                                                 
\hline
\end{tabular}
\end{center}
\end{table}

\begin{table}
\caption{Differences in the mean values of $a$, $e$, $i$  between the Grav. model and Yark. model after 1 My.}
\label{yarko_influence_forward}
\small\addtolength{\tabcolsep}{-2pt}
\begin{center}
\begin{tabular}{|c|r|r|r|}
\hline
Name            & ${\Delta} a$[AU] & \multicolumn{1}{c}{${\Delta} e$} & ${\Delta} i$[$\degr$] \\  
\hline
(20461)Dioretsa  &   +2.67e+00 &  -0.094 &  +2.900 \\
(65407)          &   -1.81e+01 &  -0.010 &  -2.360 \\
(330759)         &   +1.78e+00 &  +0.017 &  +0.231 \\
(336756)         &   -5.09e+00 &  -0.000 &  +0.090 \\
(342842)         &   -4.86e+00 &  +0.005 &  +4.961 \\
(343158)(2009 HC82) &   -1.21e-02 &  -0.016 &  -1.291 \\
(459870)(2014 AT28) &   +4.46e+01 &  +0.088 &  -2.303 \\
1999 LE31         &   -3.60e-01 &  -0.003 &  -4.758 \\
2000 HE46         &   -4.35e+01 &  -0.010 &  -1.412 \\
2004 NN8          &   -5.46e+01 &  -0.043 &  -2.347 \\
(434620)(2005 VD)   &   +6.54e+00 &  -0.040 &  +2.747 \\
2006 BZ8          &   -2.38e+01 &  +0.032 &  -1.785 \\
2008 KV42         &   +3.47e-01 &  +0.001 &  -0.161 \\
2009 QY6          &   +3.04e+00 &  +0.024 &  -1.405 \\
2009 YS6          &   +1.89e+01 &  +0.013 &  +5.832 \\
2010 CG55         &   -2.53e+01 &  -0.019 &  +0.095 \\
2010 GW64         &   +1.90e+01 &  +0.014 &  -3.312 \\
2010 OR1          &   +4.94e-01 &  +0.004 &  -1.026 \\
2011 MM4          &   +2.44e+01 &  -0.005 &  -0.388 \\
2012 HD2          &   -1.86e+01 &  -0.007 &  +0.766 \\
2013 BL76         &   +4.08e+02 &  +0.001 &  +1.888 \\
2013 LD16         &   +2.49e+01 &  +0.000 &  +1.063 \\
2013 LU28         &   +1.65e+01 &  +0.007 &  -0.921 \\
2013 NS11         &   +1.93e+02 &  -0.002 &  -1.539 \\
2014 JJ57         &   -2.47e+02 &  -0.001 &  +2.291 \\                                                 
\hline
\end{tabular}
\end{center}
\end{table}

\subsection{Dynamical lifetimes}

\begin{figure}
	\includegraphics[width=\columnwidth]{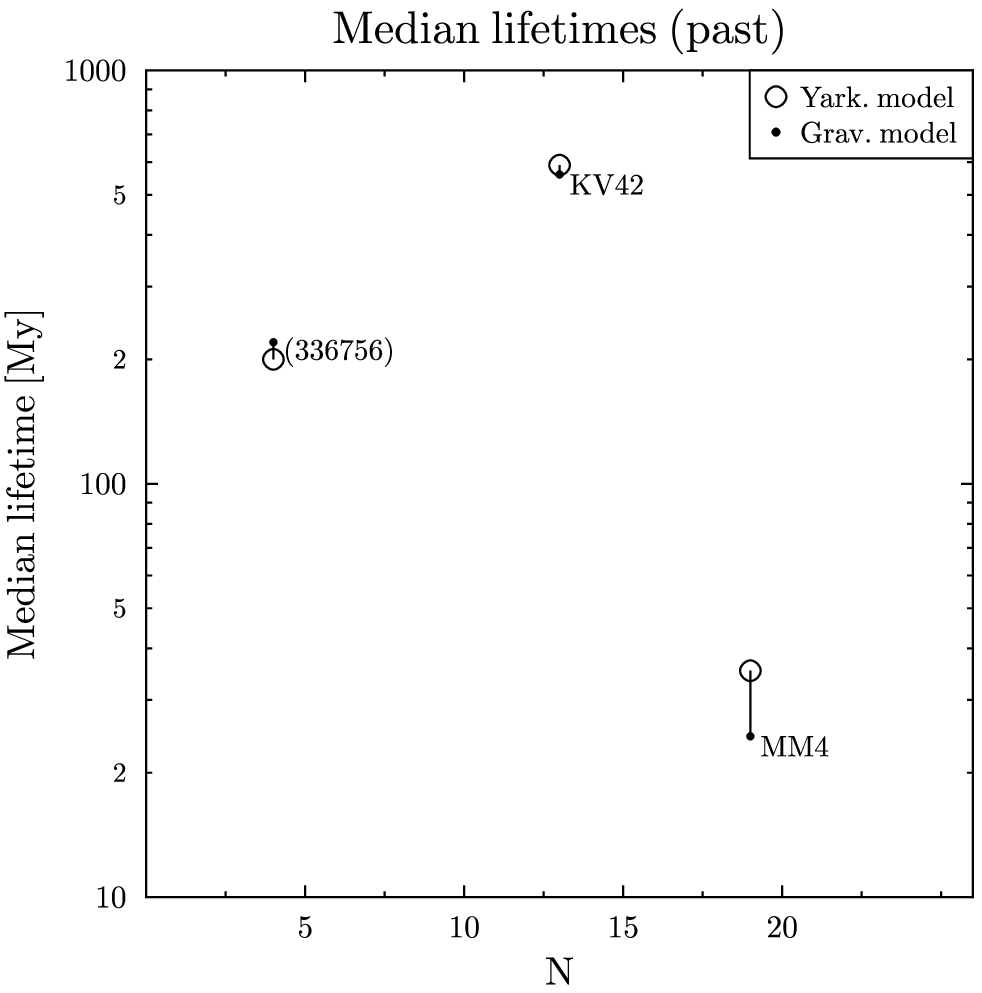}
    \caption{Median lifetimes of asteroids in retrograde orbits, obtained by numerical integration backward in time (3 objects with $\tau > 10$ My).}
    \label{lifetime_long1}
\end{figure}

\begin{figure}
	\includegraphics[width=\columnwidth]{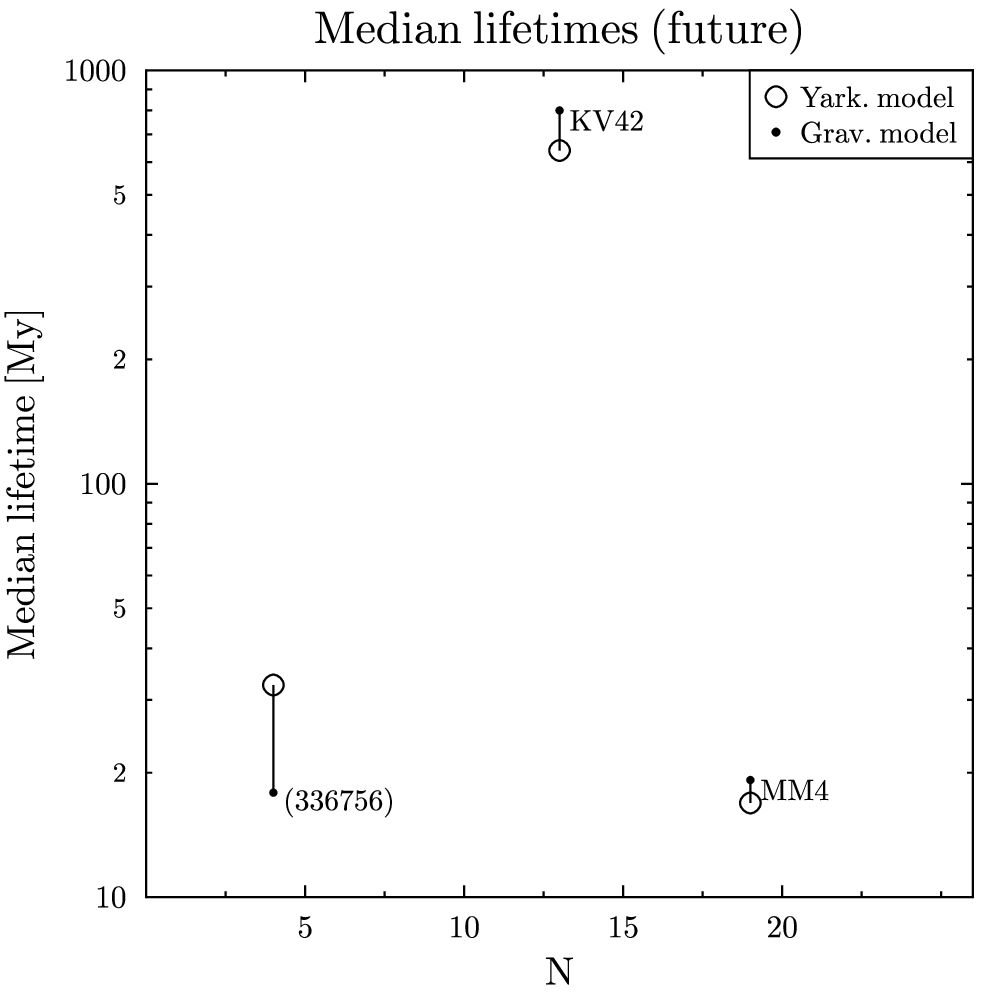}
    \caption{Median lifetimes of asteroids in retrograde orbits, obtained by  numerical integration forward in time (3 objects with $\tau > 10$ My).}
    \label{lifetime_long2}
\end{figure}

During our long-term numerical integrations, we found that test particles were ejected onto hyperbolic trajectories and removed. In many cases, the swarm of 100 clones  plus the nominal orbit did not survive the predefined time of integration ($10^8$ y). We  quantified the decay of these clones as the gradual decrease of the percentage of active test particles.
In this way we estimated the median lifetime ($\tau$) spent by the active clones on elliptical ($0<e<1$) orbits (the limit was a 50\% decay of the clones). Most  values of $\tau$ are shorter than $5$ My (Table~\ref{median_lifetimes}) and we found only three objects with  a $\tau > 10$ My (Figs~\ref{lifetime_long1} and \ref{lifetime_long2}). 
Although all three objects produced stable solutions, the most stable of them came from 2008 KV42, which exhibited a median lifetime of $\tau > 100$ My during  both backward and forward numerical  integrations. (Figs~\ref{lifetime_long1} and \ref{lifetime_long2}). For 2008 KV42, we obtained  a survival rate of 29\% (Grav. model), 34\% (Yark. model) for 1 Gy in the past, and 39\% (Grav. model), 35\% (Yark. model) for 1 Gy in the future. In the case of 2011 MM4, we obtained < 1\% (Grav. model), 5\% (Yark. model) in the past and 0\% (Grav. model), < 1\% (Yark. model) in the future. In comparison, \cite{Chen2016} obtained similar clone survival rates from the results of their numerical integrations of 2008 KV42 and 2011 MM4 (38\% and < 0.5\%, respectively).  Although some differences in our assumptions, methods, and dynamical models  can be taken into account,  we believe that these estimates are in good agreement.

The values obtained for $\tau$ are smaller than the typical median lifetimes of Centaurs with small inclinations ($i \leq 20 \degr$) \citep{Volk2013}.

\begin{figure}
	\includegraphics[width=\columnwidth]{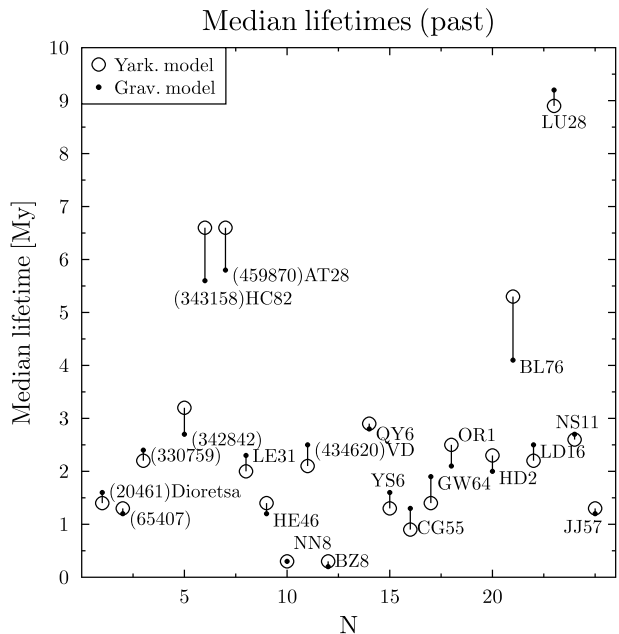}
    \caption{Median lifetimes of asteroids in retrograde orbits, obtained by numerical integration backward in time. Results obtained by Yark. model are marked by circles, while those of the Grav. model are marked by points. For each object, circles and points are linked by line segments to  highlight the difference between the results. To improve readability, we omitted the year in asteroid names.}
    \label{lifetime_figure1}
\end{figure}

\begin{figure}
	\includegraphics[width=\columnwidth]{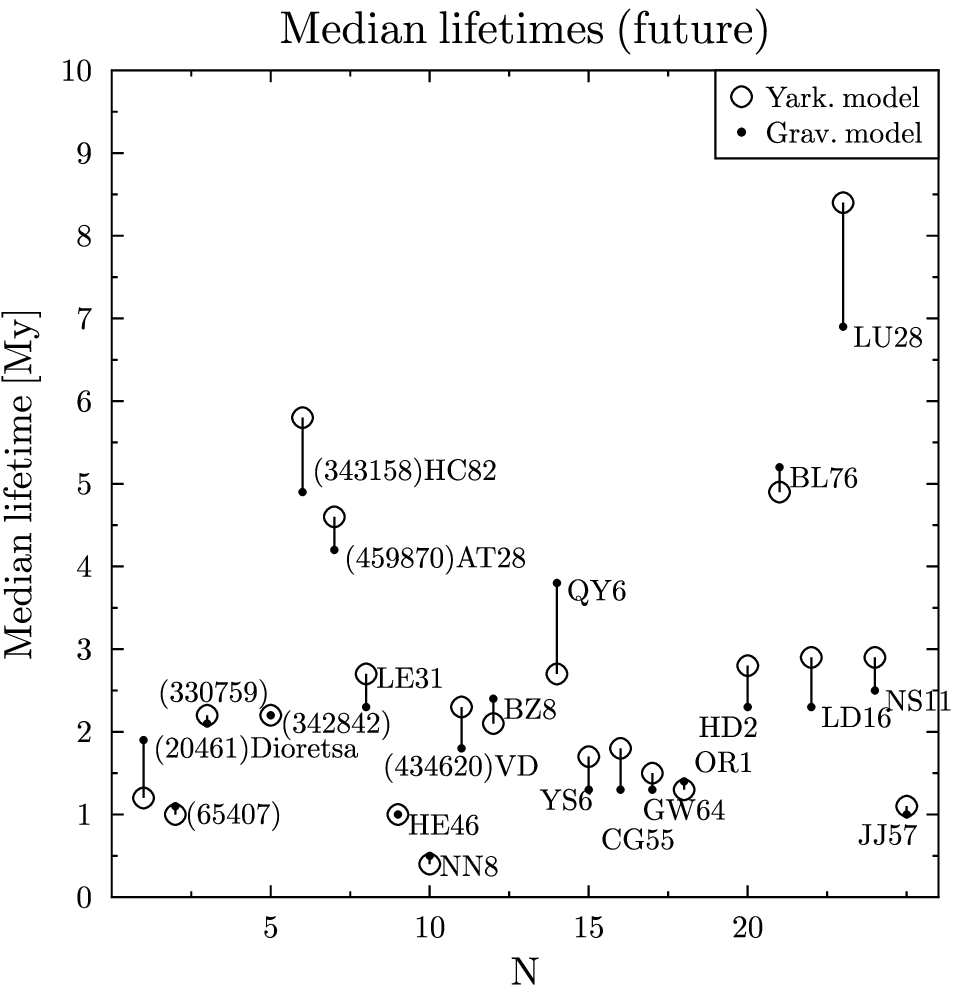}
    \caption{Median lifetimes of asteroids in retrograde orbits, obtained by numerical integration forward in time.}
    \label{lifetime_figure2}
\end{figure}

In our sample of 25 objects, $\tau$ depends on whether the integration is in the past or the future. 
The absolute values of $\tau$ are higher in the past than in the future (Table \ref{median_lifetimes}). Although the number of objects is  probably affected by small number statistics, it is possible that this is a systematic effect. Moreover, if we omit the 3 asteroids with the longest lifetimes (336756, 2008 KV42 and 2011 MM4), the average values of $\tau$ are still higher by 2.7\% to 6.5\% in the past (depending on  the model). This is visible in Figures \ref{lifetime_figure1} and \ref{lifetime_figure2}. In most cases, the lifetime is shorter in the forward integrations.

Additionally, we compared  the results obtained by the two dynamical models.  While the model of Yarkovsky forces is based on typical thermal parameters,  its influence on $\tau$ is visible as well (Figs~\ref{lifetime_figure1} \& \ref{lifetime_figure2} and Table~\ref{median_lifetimes}).  Generally, $\tau$ values are greater for the Yarkovsky model.

\begin{table}
\caption{Median lifetimes of asteroids in retrograde orbits, estimated by numerical integration with Yarkovsky forces (Yark. model) and
gravitational model.}
\label{median_lifetimes}
\small\addtolength{\tabcolsep}{-5pt}
\begin{tabular}{|c|c|c|c|c|}
\hline
Name & $\tau_{Yark}$ [My] & $\tau_{Grav}$ [My]  & $\tau_{Yark}$ [My] & $\tau_{Grav}$ [My]  \\
     & \multicolumn{2}{c}{(past)} & \multicolumn{2}{c}{(future)} \\
\hline
  (20461)Dioretsa  &  -1.4  &   -1.6   &     1.2  &  1.9 \\
  (65407)          &  -1.3  &   -1.2   &     1.0  &  1.1  \\
  (330759)         &  -2.2  &   -2.4   &     2.2  &  2.1  \\
  (336756)         & -200  &   -220 &       32.6  &  17.9  \\
  (342842)         & -3.2  &   -2.7   &     2.2  &  2.2  \\
  (343158)(2009 HC82) & -6.6  &   -5.6   &     5.8  &  4.9  \\
  (459870)(2014 AT28) & -6.6  &   -5.8   &     4.6  &  4.2  \\
  1999 LE31         & -2.0  &   -2.3   &     2.7  &  2.3  \\
  2000 HE46         & -1.4  &   -1.2   &     1.0  &  1.0  \\
 2004 NN8           & -0.3  &   -0.3    &     0.4  &  0.5  \\
 (434620)2005 VD    & -2.1  &   -2.5   &     2.3  &  1.8  \\
 2006 BZ8           & -0.3  &   -0.2   &     2.1  &  2.4  \\
 2008 KV42          & -590  &   -560  &     640 &  800 \\
 2009 QY6           & -2.9 &   -2.8   &     2.7  &  3.8  \\
 2009 YS6           & -1.3  &   -1.6   &     1.7  &  1.3  \\
 2010 CG55          & -0.9  &   -1.3   &     1.8  &  1.3  \\
 2010 GW64          & -1.4  &   -1.9   &     1.5  &  1.3  \\
 2010 OR1           & -2.5  &   -2.1   &     1.3  &  1.4  \\
 2011 MM4           & -35.3  &  -24.5  &    16.9  &  19.2  \\
 2012 HD2           & -2.3  &   -2.0   &     2.8  &  2.3  \\
 2013 BL76          & -5.3  &   -4.1   &     4.9  &  5.2  \\
 2013 LD16          & -2.2  &   -2.5   &     2.9  &  2.3  \\
 2013 LU28          & -8.9  &   -9.2   &     8.4  &  6.9  \\
 2013 NS11          & -2.6  &   -2.7   &     2.9  &  2.5  \\
 2014 JJ57          & -1.3  &   -1.2   &     1.1  &  1.0     \\                                                  
\hline
\end{tabular}
\end{table}

\section{Conclusions}
In this study, we found that asteroids in retrograde orbits are generally unstable. Many of the test particles were ejected from the Solar System during integration  and most asteroids did not survive longer than a few  million years.

The results of integrations employing two models allow for the   study of the possible influence the Yarkovsky effect has on retrograde asteroids.  
This force can potentially play an important role, especially for time-scales of $10^6$ y  where its presence could even slightly lengthen the dynamical lifetimes of these objects.
Using a simple model of thermal accelerations, we have shown that the influence of the Yarkovsky effect on most retrograde orbits can be more significant in the past than in the future. Particularly, the values of the average change of inclination, ${\Delta} i$, appear to be more sensitive to the choice of model when estimated for the past (Table~\ref{yarko_influence_backward}). Any scenarios of 'orbit inversion' or other rapid changes of $i$ in the past should take Yarkovsky forces into account. Similar behaviour is   observed for ${\Delta} e$ values. After comparing the results of the backward and forward integration, we  conclude that rapid eccentricity changes were more evident in the past.  The possibility of this occurrence can be explained by the fact that asteroids pass through different dynamical regimes in the past than in the future. To more accurately explain this phenomenon, we need a better   constraints on the thermal parameters used in Yark. model.

Unfortunately, thermal properties have not yet been sufficiently determined. We still require more precise information about the spins (pole coordinates) and periods of rotation of most retrograde objects. Currently, our results show a mean influence of the Yarkovsky effect, which was determined by the averaging. 
We expect an improvement of the dynamical model after new observations and a better determination of the thermal parameters  in the near future. This in turn should also refine the results of the long-term integrations. Additionally, corrections should be made to the dynamical models of a few retrograde minor bodies if these objects are found to exhibit cometary activity in the future by including the additional influence of non-gravitational forces, typical for comets.

Of all the objects we   have studied, only three of them, (336756), 2008 KV42, and 2011 MM4,   exhibited long median lifetimes (> 10 My),  with orbits appearing  more stable than others. For the remaining 22 asteroids, the values of $\tau$ are on the order of $\sim$ 1 My in the past and  the future. Of these 22 asteroids, 2004 N88 produced the shortest lifetimes: $-0.3$ My and $0.4$ My. In comparison with most asteroids, the lifetimes of retrograde objects in the Solar System are short.
We believe that the results presented in this paper may  be useful for studies on the possible migration of minor bodies onto retrograde orbits.

\section*{Acknowledgements}

The authors wish to thank the anonymous reviewer for many helpful suggestions and corrections relating to this article.



\bibliographystyle{mnras}
\bibliography{bibliography_pkiw} 








\bsp	
\label{lastpage}
\end{document}